# The Influences of Pre-birth Factors in Early Assessment of Child Mortality using Machine Learning Techniques

Asadullah Hill Galib[a]*, Nadia Nahar[a], and B M Mainul Hossain[a]

[a]*Institute of Information Technology, University of Dhaka, Dhaka, Bangladesh;*

*Email: bsse0712@iit.du.ac.bd; Address: Shopno Nir Tower, 52/3-C, Balughat, Dhaka Cantonment, Dhaka – 1206, Bangladesh.

# The Influences of Pre-birth Factors in Early Assessment of Child Mortality using Machine Learning Techniques

Analysis of child mortality is crucial as it pertains to the policy and programs of a country. The early assessment of patterns and trends in causes of child mortality help decision-makers assess needs, prioritize interventions, and monitor progress. Post-birth factors of the child, such as real-time clinical data, health data of the child, etc. are frequently used in child mortality study. However, in the early assessment of child mortality, pre-birth factors would be more practical and beneficial than the post-birth factors. This study aims at incorporating pre-birth factors, such as birth history, maternal history, reproduction history, socio-economic condition, etc. for classifying child mortality. To assess the relative importance of the features, Information Gain (IG) attribute evaluator is employed. For classifying child mortality, four machine learning algorithms are evaluated. Results show that the proposed approach achieved an AUC score of 0.947 in classifying child mortality which outperformed the clinical standards. In terms of accuracy, precision, recall, and f-1 score, the results are also notable and uniform. In developing countries like Bangladesh, the early assessment of child mortality using pre-birth factors would be effective and feasible as it avoids the uncertainty of the post-birth factors.

**1. Introduction**

Despite the child mortality rate is decreasing all over the world, it is still an alarming issue in the underdeveloped and developing countries. Child mortality is one of the most critical concerns of Bangladesh. Bangladesh's child mortality rate was 32.4 deaths per 1,000 live births in 2018 as reported by UNICEF.[1] According to the Central Intelligence Agency (CIA) of USA Bangladesh ranked 61[th] among 225 countries in infant mortality.[2] In global context, 15,000 children under-5 die every day all over the world.[3] That immense number of children die due to a lack of facilities and many other factors.

Factors and trends associated with child mortality might also help in classifying child mortality. According to existing works, post-birth factors, such as clinical and health-related real-time factors are crucial in child mortality analysis.[4-8] However, the incorporation of pre-birth factors, such as birth history, socio-economic attributes, family history, maternal history, etc. could be beneficial in the early classification of child mortality. Moreover, trends in the causes of child mortality serve as important global health information to guide efforts to improve child survival. Besides, trends and factors associated with child mortality differ from country to country due to living standards, socio-economic conditions, environmental factors, etc. For instance, Hobcraft et al. suggests that mother education levels tend to be strongly correlated with child mortality in Asian countries.[9] So, identifying those trends and factors associated with child mortality, utilizing those in classifying child mortality would be helpful to prevent child death.

In developing countries like Bangladesh, there are several studies that try to sort out trends and factors associated with child mortality. With the child mortality declining in Bangladesh, the distribution of causes of death also changes. Liu et al. investigated the trends in causes of death among children under 5 in Bangladesh.[10] Accordingly, Pneumonia remained the top killer of children under 5 in Bangladesh between 1993 and 2004. Many other factors and trends regarding child mortality in Bangladesh are discussed in the related literature, such as sex of the child, mother's and father's education, mother's age, preceding birth interval, place of residence , mother work status, place of delivery, multiple births, breastfeeding, birth age, period of birth, age-specific effects of previous child's death  etc.[11-18] However, to the best of our knowledge, no study deals with child mortality classification or prediction in the context of Bangladesh. Apart from that, most of the relevant studies regarding child mortality prediction or classification incorporated many post-birth features, such as clinical features, health features, and many other real-time data of the child.[4-8] Though incorporating pre-birth features might be decisive in the early assessment of child mortality classification, those features are mostly ignored in the related works.

This study aims to analyze the feature importance in classifying child mortality and incorporate pre-birth factors in classifying child mortality in Bangladesh. By doing so, a crucial healthcare issue – child mortality can be assessed earlier. Dataset from Bangladesh Demographic and Health Survey (BDHS) is used to extract relevant features from birth history, maternity, reproduction, and socio-economic information. According to existing works, feature selection takes into account the trends and factors associated with child mortality in Bangladesh. With the selected features, the feature importance is analyzed by measuring the Information Gain of each feature. Here, the Information Gain Attribute Evaluator with Ranker Search Method are used for measuring relative feature importance. Consequently, a ranking of features is generated using the features according to relevant importance in classifying child mortality. Next, for the classification of child mortality, the selected features are used to model machine learning classifiers. The model is trained using the past data records of both successful survival and child death in Bangladesh. Four machine learning classifiers are used for model training - Support Vector Machine (SVM), Random Forest, Logistic Regression, and Naive Bayes. The comparative performance of each model is evaluated by running each model through 10-fold cross-validation.

To evaluate the performance of child mortality classification, different metrics, such as AUC (Area Under the receiver operating characteristic Curve), precision, recall, f-1 scores, and accuracy is measured. According to existing works, AUC is mostly used and suitable metric for performance evaluation. The results show that the performance of the child mortality classification in terms of AUC scores is high in comparison with relevant studies and clinical standards.[4,20,21] The results of other metrics are also consistent and noteworthy. For Instance, Decision Tree classifier gives precision score of 0.943, recall score of 0.971, f-1 score of 0.957, and accuracy of 0.921 respectively. In classifying child mortality, Decision Tree and Random Forest outperformed other classifiers.

This study aims to answer two research questions (RQ) concerning feature importance and performance of the child mortality classification:

- How do the pre-birth factors perform in classifying child mortality?
- What is the relative feature importance in classifying child mortality?

The first question seeks to analyze the relevance and influence of pre-birth factors in classifying child mortality. The second question tries to figure out the relative significance of the factors in child mortality classification.

To the best of our knowledge, this is the first work on the classification of child mortality in Bangladesh using machine learning techniques.

**2. Related Work**

Researchers investigated many factors associated with child and infant mortality in Bangladesh. For instance, Kabir et al. analyzed several factors influencing child mortality in Bangladesh, such as mother's and father's education, type of place of residence (rural or urban), work status of mother, sex of the child, place of delivery (home or hospital), mother's age at the time of birth, etc.[12] D'Souza et al. suggested that female mortality is almost 50 percent higher than male mortality in Bangladesh.[11] Alam et al. suggested that maternal age is crucial for high neonatal mortality.[14] According to their analysis, teenage motherhood is one of the vital factors for neonatal mortality. In another work, they investigated the age-specific effects of previous child's mortality on the subsequent child's mortality.[18] Majumdar et al. used multivariate analysis to demonstrate that preceding birth interval length, followed by the survival status of the immediately preceding child, are the most significant factors for child and infant mortality.[13] Besides, they hinted that sex of the index child and mother's and father's education are also important regarding the mortality rate. According to Hong, multiple births in Bangladesh, regardless of other risk factors, are highly negatively associated with child mortality.[16] Miller et al. suggests that children born within 15 months of a previous birth are 60 to 80% more prone to child death.[15] According to Mondal et al. the most important predictors of neonatal, post-neonatal and infant mortality are immunization, breastfeeding, birth age and period of birth.[17]

Rinta-Koski et al. used post-birth factors associated with the Bayesian Gaussian process classification for predicting preterm infant in-hospital-mortality.[4] They collected sensory data from the first 24 hours at the Neonatal Intensive Care Unit (NICU). They incorporated those sensor measurements with standard clinical features, such as oxygen saturation by pulse oximetry (SpO2), supplemental oxygen levels, etc. Their predictive model gained an AUC of 0.94. Their performance exceeded the AUC of score 0.915 reported by Saria et al., AUC score of 0.913 for CRIB-II (Clinical Risk Index for Babies) and AUC score of 0.907 for SNAPPE-II (Score for Neonatal Acute Physiology with Perinatal Extension) reported by Reid et al.[20,21]

Zernikow et al. outperformed the Rinta-Koski et al. and other relevant models with an AUC score of 0.95.[22] They used an artificial neural network for the predictive model. However, their model trained on the older dataset - infants born between 1990 and 1993.

Wilson et al. made use of a Bayesian network to determine complex factors causing neonatal mortality.[5] Their network considered factors that cause neonatal mortality, including maternal death, preterm birth, breathing at birth and movement. They used observational data from the Child Health Registry and Global Network Maternal and used logistic regression models and algorithms that estimated the average causal effect and uncovered previously unknown factors and pathways.

Ramos et al. used Data Mining techniques for analysing infant mortality over the integrated SIM and SINASC databases of the municipality of Rio de Janeiro, RJ, Brazil, between the years 2008 and 2012.[6] This research used 13 features from the dataset instances, such as gender of the new-born, 1-minute Apgar score (5 parameters that are evaluated during the first minute of the child's life), heart rate, breathing, muscle tone, irritability, and the color of the skin. Other attributes are the 5-minutes Apgar score, weight, color, age of the child, the cause of the death, mother's age, number of mother's dead children, the number of mother's living children, number of weeks of gestation, type of pregnancy, and type of birth. However, in their work, the Naïve Bayes classifier presented better performance than the other Data Mining approaches. Besides, good results were obtained by the application of the non-supervised Apriori algorithm. As an important achievement of this work, some rules were found that could help health professionals in their everyday activities. Though it used many post-birth factors, it utilized some pre-birth factors also.

A study on births in Bega Obstetrics and Gynecology Clinique, in Timisoara, Romania, was presented by Robu and Holban.[7] It analyzed 2325 births based on 15 features such as mother's age, the number of gestations, number of weeks of pregnancy, child's gender, child's weight, and type of delivery. They looked for a new way to estimate the child's Apgar score at birth. To do so, they used 10 classification algorithms such as Naïve Bayes, ID3, KNN, Random Forest, SMO, AdaBoost, LogitBoost, JRipp, REPTree, and SimpleCart. After several experiments, the LogitBoost algorithm was selected as the best algorithm as it achieved an accuracy of 80.25%. Besides, they provided a Java application based on that.

Zhou et al. used Artificial neural networks to predict outcomes in a neonatal intensive care unit (NICU).[8] They used a gradient ascent method for the weight update of three-layer feed-forward neural networks. The results were evaluated by performance measurement techniques, such as the Receiver Operating Characteristic Curve and the Hosmer-Lemeshow test. The resulting models applied as mortality prognostic screening tools.

Most works incorporated post-birth factors for child mortality analysis, prediction and classification. The pre-birth factors are mostly ignored in the existing literature. Though several works dealt with individual pre-birth factors for analysing patterns and trends in Bangladesh, no work incorporated those pre-birth factors for child mortality prediction or classification.

## 3. Methodology

### 3.1 Dataset

This research derived relevant data from a nationally representative secondary data set - Bangladesh Demographic and Health Survey (BDHS), 2014, which was obtained through a collaborative effort by the National Institute of Population Research and Training (Bangladesh), ICF International (USA), and Mitra and Associates (Bangladesh), and United States Agency for International Development (USAID)/Bangladesh.[19] The 2014 BDHS collected data from the ever-married women (age 15-49) regarding a complete history of their live births, including the socio-economic condition, reproduction history, maternity history, sex, survival status, age at the time of the survey or age at death, etc. It also contains data on neonatal, post neonatal, infant, child, and under-5 mortality. There are 43772 instances.

*3.2 Data Pre-processing*

Firstly, features relevant to reproduction, birth history, and maternity history are considered for this research. Also, features regarding socioeconomic status, such as education years of mothers, wealth status, etc. are considered. Factors and trends associated with child death in Bangladesh as well as in Asian countries as mentioned in the related works are considered while selecting the features.[9,11-18] As the data came from a survey, many values of particular attributes left blank/unanswered. Those attributes were dropped off from the dataset due to a lack of instances.

According to the selection, the features are as follow:

(1) *Interval from previous birth*: The previous birth interval is measured as the difference between the last birth and the previous birth in months. It ranges from 9 to 267.
(2) *Total sons died*: It refers to the number of sons who have already died. It ranges from 0 to 10.
(3) *Total daughters died*: It refers to the number of daughters who have already died. It ranges from 0 to 5.
(4) *Wealth status*: It refers the wealth status of the mother. There are five categories - poorest, poorer, middle, richer, and richest.
(5) *Total number of births in the last five years*: It refers to all births in the last five years. It ranges from 0 to 5.

(6) *Number of family members*: It refers to the total number of family members. It ranges from 1 to 25.

(7) *Eligible female members in the family*: The number of eligible female members in the family. It ranges from 0 to 7.

(8) *Total children born*: It refers to the total number of children ever born. It ranges from 1 to 15.

(9) *Education years of mother*: It refers to the total number of education years of the mother. It ranges from 0 to 17.

(10) *Are children twin*: It refers to whether the last child is a twin or not.

(11) *Age of mother at first birth*: It refers to the age of the mother at the first birth of her. It ranges from 10 to 46.

(12) *Birth Month*: Month of birth of the last child. It ranges from 1 to 12.

(13) *Number of breastfeeding months of the last child*: It refers to the duration of breastfeeding of the last child in months. It ranges from 0 to 35.

(14) *Sex of child*: It refers to whether the sex of the child is male or female.

(15) *Is previous children cesarean*: It refers to whether the previous child born in the last three/five years was born by cesarean or not.

(16) *Size of child on birth*: Children size at birth, five categories - very small, small, average, larger than average, and smaller than average.

(17) *Is last children cesarean*: It refers to whether the child was born by the cesarian section or not.

Here, the target class is:

Has died: It refers to whether the last child was alive or dead at the time of the interview.

After the selection, normalization was applied to the numeric attributes. To shift the values of the numeric attributes in the dataset to a common scale, without distorting differences in the ranges of values, normalization was applied here.

*3.3 Feature Selection Technique*

In this study, Information Gain Attribute Evaluator is used for feature selection and measuring relative importance of features.[23,24] Information Gain (IG) is an entropy-based feature evaluation method, which is defined as the amount of knowledge gained from the features. It evaluates the worth of an attribute by measuring the information gain with respect to the class. First, it calculates the Entropy. Entropy measures the degree of "impurity". Entropy should be reduced for better classification as higher Entropy induces uncertainity. The Entropy of a set T, E(T), is defined as follows:

$$E(T) = -\sum_{i=1}^{C}(P_i * \log_2 P_i) \qquad (1)$$

Where, $C$ is the set of classes in $T$, and $P_i$ is the proportion of the number of elements in class $i$ to the number of elements in set $T$.

According to the Entropy, the Information Gain Attribute Evaluator calculates the Information Gain (IG). The Information Gain of $T$ for attribute $a$, IG (T, a) is defined as follows:

$$IG(T, a) = E(T) - E(T|a) \qquad (2)$$

Where, $E(T/a)$ is the conditional entropy of $T$ given the value of attribute $a$.

Using 10-fold cross-validation on the Weka tool, Information Gain Attribute Evaluator is employed.[24] It assesses the weight of an attribute by calculating the information gain with respect to the class. Afterwards, Ranker Search Method of Weka tool ranks attributes by their individual evaluations. Here, that evaluator finally provided a ranking of features based on their importance for classifying child mortality.

*3.4 Classification Approach*

According to the features, a machine learning model is trained for classifying child mortality. Four machine learning algorithms are evaluated here:

*Support Vector Machine (SVM)*: Support Vector Machine is a two-group classification problem learning technique. Here, the input vectors are translated to a very high dimensional feature space in a non-linear manner. A linear decision surface is built in this feature space. Different characteristics of the decision surface ensure the learning machine has a high generalization power.[25]

*Random Forest*: Random forests are an ensemble learning method that works by building a variety of decision trees during training. The purpose of the approach is to create multiple trees in the feature space randomly selected subspaces. Trees in various subspaces generalize their classification in complementary ways and can monotonically boost their combined classification.[26]

*Logistic Regression*: Logistic regression is a statistical model which employs a logistic function to model a binary dependent variable in its fundamental form. When the dependent variable is categorical, logical regression is used mostly. Several medical scales were established using logistic regression to measure the severity of a patient.

*Naive Bayes*: The classification of Naive Bayes is a set of classification algorithms based on the Bayes theorem. Each pair of features to be classified is independent from each other. It is employed in automatic medical diagnosis.

All these machine learning models are popular and well-suited for different cases. These models were fitted with the training data and cross-validated in 10 folds using the test data. The parameters of the different machine learning models are set to default in Weka tool as that configuration gives high performance values.

By running each model through 10-fold cross-validation, the relative performance of each model is assessed. For each of the 10 repetitions of the analysis, the dataset is split into 10 equal partitions, commonly referred to as folds, assuring that instances, where *Has died* is true, are distributed equally across folds. This follows generally accepted benchmarks for accurate training and testing of model performance.[27]

## 4. Findings and Result Analysis

*4.1 Feature Importance (Ranking)*

The ranked features set according to the Information Gain Attribute Evaluator and Ranker Search Method are given in Table 1.

According to the feature ranking, total sons and daughters died are the most important features for classifying child mortality. Apart from the birth and maternity history-related features, particular socio-economic features, such as education years of mothers, and the wealth status of the family are notable features in classifying child mortality. Also, features like the number of members and eligible female members in the family, age of the mother at first birth, the interval from previous birth play a decisive role in child mortality classification. These results suggest that pre-birth factors like birth history, maternity history, socio-economic condition, etc. have consequential impact on child mortality analysis.

*4.2 Classification Performance*

According to relevant literature, to measure whether the model gives higher probabilities to children with the event (e.g. children who die) as opposed to children without the event, the AUC (Area Under the receiver operating characteristic Curve) is most commonly used. So, in evaluating the performance of the model, AUC is considered primarily. Also, other metrics for model evaluation, such as accuracy, precision, recall, and F1-score are calculated in this study. Four machine learning models (SVM, Random Forest, Logistic Regression, Naïve Bayes) are evaluated in this study. The results of the evaluation are depicted in Table 2.

In terms of AUC, Decision Tree and Random Forest have high AUC scores of 0.947 and 0.940 respectively. Though this study is in the context of Bangladesh only, it exceeds the performance of relevant studies and clinical standards, such as the AUC scores of 0.94 (Rinta-Koski et al.), 0.915 (Saria et al.), 0.913 and 0.907 (CRIB-II and SNAPPE-II, reported by Reid et al.) respectively.[4,20,21] The high AUC score of this study suggests that the pre-birth factors are crucial for classifying child mortality.

The classification problem in this study is life-critical as identifying a birth at high risk is more important than misclassification of a low-risk birth. So, in this case, the recall score is an important metric as the results should contain less False Negatives to give an effective classification. According to Table 2, all the models using different algorithms give high recall scores ranging from 0.951 to 0.981.

In addition, the precision scores range from 0.924 to 0.943, F-1 scores range from 0.943 to 0.953, and accuracy range from 0.895 to 0.921 for all the four machine learning algorithms. These scores help to solidify the base of pre-birth factors in classifying child mortality. Moreover, it should be noted that all of these performance values do not fluctuate notably for different models, which might suggest that the features are well suited for child mortality classification.

From Table 2, it is clear that Decision Tree dominates over the other algorithms in terms of precision, f-1 score, and accuracy. Also, Random Forest and Logistic

Regression perform comparatively well with respect to Naive Bayes. It might be suggested that Decision Tree can be used in the current scenario for achieving high performance.

**5. Limitation**

Regarding threats to internal validity, the results of this study are difficult to interpret due to methodological complexity. The feature importance from certain methods does not imply direction, making it more difficult to infer relationships. Besides, the 10-repetition, 10-fold cross-validation framework requires variable importance and inclusion to be aggregated across run and fold combinations. However, this complexity allows for a much deeper examination and cross-validation of the factors that influence in-hospital mortality in this case. Additionally, extended parameter tuning could not be performed on several methods due to computational limitations. Although tuning the parameters of each model could improve their performance, the baseline models performed relatively well, and tests of parameter tuning resulted in minimal performance gain and significant computational strain.

In terms of threats to external validity, the generalizability of the results might be questionable as this study uses a national dataset in the context of Bangladesh only. So, its potential to be applied to other contexts should be reassessed. It might be possible that factors associated child mortality varies from region to region.

Therefore, the proposed approach and features set have to be well-analyzed and should not be used in performing real-time risk prediction before further analysis. As child mortality is a life-critical issue, the study needs more reviews and analytical advancement to be a generalized solution in practice.

## 6. Conclusion

Identifying significant factors associated with child mortality would be beneficial for building an advanced prevention strategy. This research seeks to sort out the factors associated with child mortality and their relative significance. Also, it directs at the applicability of many pre-birth factors in classifying child mortality in developing countries like Bangladesh which might be beneficial for preventing this phenomenon. This is the first study in classifying child mortality in Bangladesh and considering pre-birth factors for classification according to the best of our knowledge.

While existing works of child mortality analysis mostly dealt with post-birth factors in classifying child mortality, this study also incorporated pre-birth factors in classifying child mortality. Results show that incorporating pre-birth factors (birth history, maternity, reproduction, and socio-economic information, etc.) performs better in terms of AUC scores in comparison with relevant studies and clinical standards in classifying child mortality.[4,20,21]

This study also includes the identification of a statistically significant pattern: feature importance and ranking in classifying child mortality in Bangladesh. That ranking could be used for targeted investment of resources to mitigate the risks. It also suggests that child mortality depends on these ranked features mentioned in this study as these features give sound performance in child mortality classification. So, taking the necessary steps regarding these features may reduce child mortality. Besides, further surveys and finding new factors should be maintained. Finally, this work can be helpful for the government and hospital to take an insight on this burning issue and prioritize activities regarding child mortality reduction.

Future works will involve the study of other decisive factors, such as psychological factors, social factors, demographic factors, etc. for child mortality classification. Also, incorporating physical and health information of children in classifying child mortality will be performed in the future.

## 7. Disclosure of Interest

No potential conflict of interest was reported by the authors.

## 8. Ethical Consideration Statement

Not applicable. As part of this work, no experiments were conducted on humans or animals.

Table 1. Ranking of Features based on Information Gain.

| Ranking | Features |
|---------|----------|
| 1 | Total sons died |
| 2 | Total daughters |
| 3 | Total children born |
| 4 | Education years of mother |
| 5 | Are children twin |
| 6 | Interval from previous birth |
| 7 | Wealth status |
| 8 | Total number of births in the last five years |
| 9 | Number of family members |
| 10 | Age of mother at first birth |
| 11 | Eligible female members in family |
| 12 | Birth Month |
| 13 | Number of breastfeeding months of the last child |
| 14 | Sex of child |
| 15 | Is previous children caesarean |
| 16 | Size of child on birth |
| 17 | Is last children caesarean |

Table 2. Classification Performance

|  | AUC | Precision | Recall | F-1 Score | Accuracy |
|---|---|---|---|---|---|
| Decision Tree | 0.947 | 0.943 | 0.971 | 0.957 | 0.921 |
| Random Forest | 0.940 | 0.940 | 0.956 | 0.948 | 0.905 |
| Logistic Regression | 0.916 | 0.924 | 0.981 | 0.953 | 0.912 |
| Naïve Bayes | 0.861 | 0.935 | 0.951 | 0.943 | 0.895 |